\documentclass[12pt]{article}
\usepackage{amsmath}
\usepackage{color}
\usepackage{graphicx}
\begin{document}
\baselineskip=18 pt
\begin{center}
{\large{\bf The generalized Klein-Gordon oscillator in the background of cosmic string space-time with a linear  potential in the Kaluza-Klein theory }}
\end{center}

\vspace{.5cm}

\begin{center}
{\bf Faizuddin Ahmed}\footnote{faizuddinahmed15@gmail.com ; faiz4U.enter@rediffmail.com}\\ 
{\bf Ajmal College of Arts and Science, Dhubri-783324, Assam, India}
\end{center}

\vspace{.5cm}

\begin{abstract}

In this work, we study the generalized Klein-Gordon oscillator with interactions on a curved background within the Kaluza-Klein theory. We solve the generalized Klein-Gordon oscillator in the cosmic string space-time with a linear scalar potential and obtain the energy eigenvalue and corresponding eigenfunction. We show that the energy spectrum depends on the global parameters characterizing the space-time and the confining potential parameter. We also solve the generalized Klein-Gordon oscillator in a magnetic cosmic string background in the Kaluza-Klein theory with a linear scalar potential and analyze the analogue effect to the Aharonov-Bohm effect for bound states.

\end{abstract}

{\bf keywords:} Klein-Gordon oscillator, Topological defects, Kaluza-Klein theory, confining potentials, relativistic wave equaions : bound state solutions, special functions : Nikiforov-Uvarov method, Heun's differential equation.

\vspace{0.1cm}

{\bf PACS Number(s):} 03.65.Pm, 03.65.Ge, 04.50.Cd, 04.50.+h, 11.10.Kk

\section{Introduction}

The relativistic quantum dynamics of scalar and spin-$\frac{1}{2}$ particles on curved background space-time geometries as well as G\"{o}del, and G\"{o}del-type metrics have been investigated by various authors (see \cite{aa1} and references therein). The Klein-Gordon and Dirac equations in a G\"{o}del-type space-times with positive, zero and negative curvatures were first studied in \cite{aa2}. The close relationship between the quantum dynamics of the scalar particle in the background of general relativity with the G\"{o}del solutions and the Landau levels in flat, spherical and hyperbolic spaces were investigated in \cite{aa3}. Later, the same problem was studied by solving the Klein-Gordon equation in the Som-Raychaudhuri space-time in \cite{aa4}. The authors in \cite{aa5} solved the Klein-Gordon equation in a family of G\"{o}del-type solutions with the cosmic string and analyzed the similarity of the energy eigenvalue with the Landau levels in flat, spherical and hyperbolic spaces. Quantum influence of topological defects in a G\"{o}del-type space-times in flat, spherical and hyperbolic cases, were investigated in \cite{aa6}. The relativistic quantum dynamics of a Dirac particle with topological defects in a G\"{o}del-type space-times with torsion have been investigated in \cite{aa7}. The relativistic quantum dynamics of an electrically charged particle described by the Klein-Gordon oscillator subject to a Coulomb-type potential was investigated in \cite{cc5}. Weyl fermions in a family of G\"{o}del-type geometries with topological defects were investigated in \cite{aa8}. The relativistic quantum dynamics of a scalar particle in 4D curved space-time with the cosmic string was investigated in \cite{EPJP}. The relativistic quantum dynamics of scalar and spin-$\frac{1}{2}$ particles subject to various kind of potentials have been investigated in several areas of physics ({\it e. g.}, \cite{aa9,aa10,aa11,aa12,aa13,aa14,aa15,aa16,aa17,aa18,aa19,aa20,aa21,aa22,aa23}). Linear confinement of quantum particle by introducing a linear scalar potential into the relativistic system by modifying the mass term has great importance for models of confinement of quarks \cite{Chric}. It is worth mentioning that the linear scalar potential has attracted a great interest in atomic and molecular physics \cite{aa25,aa26,aa27,aa28,aa29,aa30}, and in the relativistic quantum mechanics \cite{EPJC,EPJC2,d12,d16,d15,aa31,aa32, aa33,aa35,aa36,aa37,aa38,aa40,aa47,aa48,aa49, aa50,aa51}.

Interactions of the Dirac oscillator with the gravitational fields produced by topological defects were investigated in \cite{cc1}. The influence of Aharonov-Casher effect on the Dirac oscillator in three different scenarios of general relativity: the Minkowski space-time, the cosmic string and the cosmic dislocation space-time were studied in \cite{cc2}. The influence of non-inertial effects on the Dirac oscillator in the cosmic string space-time was investigated in \cite{cc3}. The Dirac equation in a class of topologically trivial flat G\"{o}del-type space-time was investigated in \cite{EPJC3}. Dirac fermions in the Som-Raychaudhuri space-time with a linear scalar and vector potentials were investigated in \cite{d12}. A new model for study the confinement of spin-half particles in a two-dimensional quantum ring systems described by the Dirac equation with a new coupling were studied in \cite{cc4}. The Dirac oscillator in the context of Doubly General Relativity was investigated in \cite{cc6}. Effects of gravitational fields produced by topological defects on the Dirac field and oscillator in a spinning cosmic string was examined in \cite{cc7}. The dynamics of 2D Dirac oscillator in the space-time of a magnetic cosmic string were investigated in \cite{cc8}. The generalized Dirac oscillator in the cosmic string space-time replacing the momentum $p_{\mu}$ with its alternative $p_{\mu}+m\,\omega\,\beta\,f_{\mu} (x_{\mu})$ was studied in \cite{cc9}. In particular, the quantum dynamics was considered for the function $f_{\mu} (x_{\mu})$ to be taken as Cornell-type, exponential-type and singular potentials form. The generalized Dirac oscillator was introduced in (2+1)-dimensional the world \cite{cc10}. The Dirac oscillator under the influence of non-inertial effects in a rotating frame in the cosmic string space-time were investigated in \cite{cc11}. The Dirac oscillator has also been analyzed in various physical systems, such as in the presence of external fields \cite{cc12}, and in the presence of a magnetic quantum flux \cite{cc13,cc14,cc15,cc16}. Investigation of magnetization and persistent current of mass-less Dirac fermions confined in a quantum dot in a graphene layer with topological defects were done in \cite{cc17}. Non-inertial effects on the Dirac oscillator in the background space-time generated by a cosmic string have been investigated \cite{cc18,cc19,cc20}. The (1+2)-dimensional Dirac oscillator in the presence of a homogeneous magnetic field in an Aharonov-Casher system were investigated in \cite{cc21}. The relativistic quantum dynamics of spin-half particle by solving the Dirac equation in (1+2)-dimensional G\"{u}rses space-time was investigated in \cite{cc22}.

The Klein-Gordon oscillator \cite{bb1,bb2} was inspired by the Dirac oscillator \cite{bb3} applied to half-integer spin particles. The spectral distribution of energy levels and eigenfunction describing the state of a particle by solving the Klein-Gordon equation in one-dimensional version of the Minkowski space-time were studied in \cite{bb4}. The Klein-Gordon oscillator in the cosmic string space-time in the presence of external fields were studied in \cite{bb5}. The Klein-Gordon oscillator in the presence of a Coulomb-type potential was investigated by two ways : (i) by modifying the mass term $m \rightarrow m+S(r)$ \cite{bb6} and (ii) via the minimal coupling \cite{aa50} besides a linear scalar potential. The relativistic quantum effects on the Klein-Gordon oscillator with linear scalar and Coulomb-type potentials were investigated in \cite{aa49}. The Klein-Gordon oscillator has also been investigated in various physical system, such as in the background space-time generated by the cosmic string \cite{bb8}, in the background of a G\"{o}del-type space-time under the influence of gravitational fields produced by topological defects \cite{bb9}, in the background of the Som-Raychaudhuri space-time with a disclination parameter \cite{aa5}, Aharonov-Bohm effect for bound states on a scalar particle in a space-time with a spacelike dislocation \cite{bb48}, in non-commutative (NC) phase space \cite{bb10}, in (1+2)-dimensional G\"{u}rses space-time background \cite{ff4}, and in (1+2)-dimensional G\"{u}rses space-time background subject to Coulomb-type potential \cite{ff5}.

Our intention now is to extend the above studies not only to two-,three-, and four-dimensions but to consider this dynamics in general background space-time produced by topological defects using the Kaluza-Klein theory \cite{bb12,bb13,bb11}. This new proposal establishes that the electromagnetism can be introduced through an extra (compactified) dimension in the space-time where, the spatial dimension becomes five-dimensional. These sources of gravitational fields play an important role in condensed matter physics systems \cite{bb17,bb18,bb19,bb20}. Besides, the topological defects like the cosmic strings \cite{bb21}, domain walls \cite{bb22}, and global mono-pole \cite{bb23} provides a tiny relation between the effects in cosmology and gravitation. In condensed matter physics systems where, topological defects analogue to the cosmic strings appear in phase transitions in liquid crystals \cite{bb24,bb25}. In addition, the Kaluza-Klein theory has found wide applications in string theory \cite{bb26}, in the presence of torsion \cite{bb31,bb32}, fermions \cite{bb33,bb34,bb35}, and in the studies of Lorentz symmetry violation (LSV) in  \cite{bb36,bb37,bb38}. Based on these generalizations of topological defects space-time within the Kaluza-Klein theory, analogue effects to the Aharonov-Bohm effect for bound states were investigated in \cite{aa6,bb28,bb39,bb40, bb50}. The Klein-Gordon oscillator on curved background within the Kaluza-Klein theory were investigated in \cite{aa6}. Some other physical systems in the Kaluza-Klein theory were studied in \cite{bb14,bb15, bb16,bb27,bb29}

The relativistic quantum dynamics of scalar particle of mass $m$ is described by the Klein-Gordon equation :
\begin{equation}
\frac{1}{\sqrt{-g}}\,\partial_{\mu} (\sqrt{-g}\,g^{\mu\nu}\,\partial_{\nu}\,\Psi)=m^2\,\Psi,
\label{1}
\end{equation}
with $g$ is the determinant of metric tensor with $g^{\mu\nu}$ its inverse. If one introduces a scalar potential into the Klein-Gordon equation by modifying the mass term: $m \rightarrow m + S (r)$ \cite{WG} where, $S (r)$ is the scalar potential. The KG-equation becomes 
\begin{equation}
\frac{1}{\sqrt{-g}}\,\partial_{\mu} (\sqrt{-g}\,g^{\mu\nu}\,\partial_{\nu}\,\Psi)=(m + S(r))^2\,\Psi.
\label{2}
\end{equation}
To couple the Klein-Gordon field with the oscillator \cite{bb1,bb2}, the following change in the momentum operator is considered \cite{dd2} :
\begin{equation}
p_{\mu}\rightarrow p_{\mu}+i\,m\,\omega\,X_{\mu},
\label{3}
\end{equation}
where $\omega$ is the frequency of oscillatory particle and the vector $X_{\mu}=(0, r, 0, 0, 0)=r\,\hat{r}$ where, $r$ being the distance of the particle from the string. To generalize the Klein-Gordon oscillator we have replaced $\vec{r}$ by the function $f(r)$ into $X_{\mu}$ as:  
\begin{equation}
X_{\mu}=(0, f(r), 0, 0, 0)=f(r)\,\hat{r}.
\label{4}
\end{equation}
Therefore, we have the following generalized Klein-Gordon oscillator equation:
\begin{equation}
\frac{1}{\sqrt{-g}}\,(\partial_{\mu}+m\,\omega\,X_{\mu})\sqrt{-g}\,g^{\mu\nu}\,(\partial_{\nu}-m\,\omega\,X_{\nu})\,\Psi=(m+ S(r))^2\,\Psi.
\label{5}
\end{equation}

This paper comprises as follow : in {\it section 2}, we study the generalized Klein-Gordon oscillator in the cosmic string space-time in the Kaluza-Klein theory subject to a linear scalar potential; in {\it section 3}, the generalized Klein-Gordon oscillator in the background of a magnetic cosmic string in the Kaluza-Klein theory subject to a linear scalar potential, and finally the conclusion in {\it section 4}.

\section{Generalized Klein-Gordon oscillator in the cosmic string space-time with a linear scalar potential in the Kaluza-Klein theory}

The purpose of this section is to study the Klein-Gordon equation in the background space-time generated by the cosmic string within the Kaluza-Klein theory subject to a linear scalar potential. Linear scalar potential is introduced into the relativistic system by modifying the mass term discussed earlier and this kind of potential has great importance in different branch of physics mentioned in the introduction. 

The first study of topological defects in the Kaluza-Klein theory was carried out in \cite{bb14} where, a series of cylindrically symmetric solutions of Einstein and Einstein-Gauss-Bonet equations, were investigated. Various solutions of the cosmic string form in five-dimensions, such as: neutral cosmic string, cosmic dislocation, superconducting cosmic, multi-cosmic string space-time were studied there. The metric corresponding to this geometry can be written as,
\begin{equation}
ds^2=-dt^2+dr^2+\alpha^2\,r^2\,d\phi^2+dz^2+dx^2,
\label{6}
\end{equation}
where $t$ is the time coordinate, $x$ is the coordinate associated with the fifth additional dimensions and $(r, \phi, z)$ are cylindrical coordinates. These coordinates assumed the ranges $-\infty < (t, z) < \infty$, $0 \leq r < \infty$, $0 \leq \phi \leq 2\,\pi$, and $0 < x < 2\,\pi\,a$ where, $a$ is the radius of the compact dimension $x$. The $\alpha$ parameter characterizing the cosmic string and in terms of linear mass density $\mu$ is $\alpha=1-4\,\mu$ \cite{bb21}, where $G=1=c=\hbar$. Cosmology and gravitation imposes limits of $\alpha$ parameter which is restricted to $\alpha < 1$ \cite{bb21}. Moreover, in condensed matter physics systems, this restriction is free and the opposite case $\alpha > 1$, the known negative disclination \cite{bb29} can occur in several systems as those described by \cite{bb25}.

By considering the line element (\ref{6}) into the Eq. (\ref{5}), we obtain the following differential equation:
\begin{eqnarray}
&&[-\partial_{t}^2+\partial_{z}^2+\partial_{x}^2+\partial_{r}^2+\frac{1}{r}\,\partial_{r}+\frac{1}{\alpha^2\,r^2}\,\partial_{\phi}^2-m\,\omega\,(f'(r)+\frac{f (r)}{r})-m^2\,\omega^2\,f^{2} (r)\nonumber\\
&&-(m+ S(r))^2]\,\Psi=0.
\label{7}
\end{eqnarray}
The above equation is independent of $t, \phi ,z, x$, so we choose the following ansatz for the function $\Psi$
\begin{equation}
\Psi (t, r, \phi, z, x)=e^{i\,(-E\,t+l\,\phi+k\,z+q\,x)}\,\psi(r),
\label{8}
\end{equation}
where $E$ is the total energy, $l=0,\pm\,1,\pm\,2...\in {\bf Z}$ are the eigenvalue of the $z$-component of the angular momentum operator, and $k, q$ are constants.

Substituting the above ansatz into the Eq. (\ref{7}), we get the following radial wave-equation for $\psi (r)$ :
\begin{eqnarray}
&&\psi''(r)+\frac{1}{r}\,\psi'(r)+[E^2-k^2-q^2-\frac{l^2}{\alpha^2\,r^2}-m\,\omega\,(f'(r)+\frac{f (r)}{r})\nonumber\\
&&-m^2\,\omega^2\,f^{2}(r)-(m+S(r))^2]\,\psi(r)=0.
\label{9}
\end{eqnarray}

In this work, we consider a special kind of potential which has many applications including the linear confinement of quarks which we discussed in the introduction is given by
\begin{equation}
S(r)=k_{L}\,r.
\label{10}
\end{equation}
where $k_{L}$ is the linear confining parameter.

We choose the function $f(r)$ as Cornell-type potential form given by \cite{cc9,bb41}
\begin{equation}
f(r)=a\,r+\frac{b}{r}\quad,\quad a, b>0.
\label{function}
\end{equation}

Substituting the function Eq. (\ref{function}) into the Eq. (\ref{9}) and using the potential (\ref{10}), we obtain the following equation :
\begin{equation}
\psi''(r)+\frac{1}{r}\,\psi'(r)+[\lambda-\tilde{\omega}^2\,r^2-\frac{\xi^2_{s}}{r^2}-2\,m\,k_{L}\,r]\,\psi(r)=0,
\label{11}
\end{equation}
where
\begin{eqnarray}
&&\lambda=E^2-k^2-q^2-m^2-2\,m\,\omega\,a-2\,m^2\,\omega^2\,a\,b,\nonumber\\
&&\tilde{\omega}=\sqrt{m^2\,\omega^2\,a^2+k^{2}_{L}},\nonumber\\
&&\xi_{s}=\sqrt{\frac{l^2}{\alpha^2}+m^2\,\omega^2\,b^2}.
\label{12}
\end{eqnarray}
Transforming $\rho=\sqrt{\tilde{\omega}}\,r$ into the Eq. (\ref{11}), we obtain
\begin{equation}
\psi''(\rho)+\frac{1}{\rho}\,\psi'(\rho)+[\lambda_0-\rho^2-\frac{\xi^2_{s}}{\rho^2}-\theta\,\rho]\,\psi(\rho)=0,
\label{13}
\end{equation}
where
\begin{equation}
\lambda_0=\frac{\lambda}{\tilde{\omega}}\quad,\quad \theta=\frac{2\,m\,k_{L}}{\tilde{\omega}^{\frac{3}{2}}}.
\label{aa}
\end{equation}

Now, we use the appropriate boundary conditions to investigate the bound states solution in this problem. It is require that the regularity of wave-function at the origin and the normalizability at infinity. Then we proceed with
the analysis of the asymptotic behaviour of the radial eigenfunction at origin and in the infinite. These conditions are necessary since the wave-function must be well-behaved in these limit, and thus, bound states of the energy
eigenvalue for the confinement can be obtained. With these conditions, we obtain the convergence of the wave-function at the origin ($\rho \rightarrow 0$) and at infinity ($\rho \rightarrow \infty$). Let us impose the requirement that $\psi (\rho) \rightarrow 0$ both at $\rho \rightarrow 0$ and $\rho \rightarrow \infty$. Suppose the possible solution to the Eq. (\ref{13}) is
\begin{equation}
\psi (\rho)=\rho^{|\xi_{s}|}\,e^{-\frac{1}{2}\,(\theta+\rho)\,\rho}\,H (\rho),
\label{14}
\end{equation}
where $H (\rho)$ is an unknown function. Substituting the solution Eq. (\ref{14}) into the Eq. (\ref{13}), we obtain
\begin{equation}
H''(\rho)+[\frac{\gamma}{\rho}-\theta-2\,\rho]\,H'(\rho)+[-\frac{Q}{\rho}+\Theta]\,H (\rho)=0,
\label{17}
\end{equation}
where
\begin{eqnarray}
&&\gamma=1+2\,\xi_s,\nonumber\\
&&\Theta=\lambda_0+\frac{\theta^2}{4}-2\,(1+\xi_s),\nonumber\\
&&Q=\frac{\theta}{2}\,(1+2\,\xi_s).
\label{18}
\end{eqnarray}
Equation (\ref{17}) is the biconfluent Heun differential equation \cite{aa30,aa49,aa50,bb41,bb42,ff2,ff3} with $H (\rho)$ is the Heun polynomials function.

The above Eq. (\ref{17}) can be solved by the power series method \cite{bb43}
\begin{equation}
H (\rho)=\sum^{\infty}_{i=0} c_{i}\,\rho^{i}.
\label{19}
\end{equation}
Substituting Eq. (\ref{19}) into the Eq. (\ref{17}), we get the following recurrernce relation for the coefficients:
\begin{equation}
c_{n+2}=\frac{1}{(n+2)(n+1+\gamma)}\,[\{Q+\theta\,(n+1) \}\,c_{n+1}-(\Theta-2\,n)\,c_{n}].
\label{20}
\end{equation}
And the various coefficients are
\begin{equation}
c_1=\frac{Q}{\gamma}\,c_0,\quad c_2=\frac{1}{2\,(1+\gamma)}\,[(Q+\theta)\,c_{1}-\Theta\,c_{0}].
\label{21}
\end{equation}
The bound states solution to the Eq. (\ref{17}) can be obtained because there is no divergence of the wave-function both at $\rho \rightarrow 0$ and $\rho \rightarrow \infty$. As we have written the function $H (\rho)$ as a power series expansion around the origin in Eq. (\ref{19}). Thereby, the bound states solution can be achieved by imposing that the power series expansion becomes a polynomial of degree $n$. Through the recurrence relation Eq. (\ref{20}), we can see that the power series expansion becomes a polynomial of degree $n$ by imposing two conditions \cite{aa30,aa50,ff2,ff3,bb41,bb44,bb45,bb46,bb47}:
\begin{equation}
\Theta=2\,n.
\label{22}
\end{equation}
And
\begin{equation}
c_{n+1}=0,\quad (n=1,2,3,4,......).
\label{23}
\end{equation}
Using the condition $\Theta=2\,n$, we get the following eigenvalue $E_{n,l}$:
\begin{eqnarray}
&&\lambda_0+\frac{\theta^2}{4}-2\,(1+\xi_s)=2\,n\nonumber\\\Rightarrow
&&E^{2}_{n,l}=k^2+q^2+m^2+2\,m\,\omega\,a+2\,m^2\,\omega^2\,a\,b\nonumber\\
&&+2\,\tilde{\omega}\,(n+1+\sqrt{\frac{l^2}{\alpha^2}+m^2\,\omega^2\,b^2})-\frac{m^2\,k^{2}_{L}}{\tilde{\omega}^2}.
\label{24}
\end{eqnarray}
Equation (\ref{24}) is the bound states energy eigenvalue of a scalar particle associated with $n^{th}$ radial modes.

The corresponding wave-function is given by
\begin{equation}
\psi_{n,l} (\rho)=\rho^{|\xi_s|}\,e^{-\frac{1}{2}\,[\frac{2\,m\,k_{L}}{\tilde{\omega}^{\frac{3}{2}}}+\rho]\,\rho}\,H (\rho).
\label{25}
\end{equation}

Now, we impose the additional condition $c_{n+1}=0$ to find the individual energy levels and corresponding wave function one by one as done in \cite{aa30,aa50,bb44,bb45}. As example, for $n=1$, we have $c_2=0$ which implies
\begin{eqnarray}
&&c_1=\frac{2}{Q+\theta}\,c_0 \Rightarrow \frac{Q}{\gamma}=\frac{2}{Q+\theta}\nonumber\\\Rightarrow
&&\tilde{\omega}_{1,l}=[m^2\,k^2_{L 1}\,(\frac{3}{2}+\sqrt{\frac{l^2}{\alpha^2}+m^2\,\omega^2\,b^2})]^{\frac{1}{3}},
\label{26}
\end{eqnarray}
where we have adjusted the parameter $k_{L}$ such that the first degree polynomial solution for $n=1$ can be obtained \cite{bb41}. Therefore, the ground state energy eigenvalue for $n=1$ is given by
\begin{eqnarray}
&&E^{2}_{1,l}=k^2+q^2+m^2+2\,m\,\omega\,a+2\,m^2\,\omega^2\,a\,b\nonumber\\
&&+2\,\tilde{\omega}_{1,l}\,(n+1+\sqrt{\frac{l^2}{\alpha^2}+m^2\,\omega^2\,b^2})-\frac{m^2\,k^{2}_{L 1}}{\tilde{\omega}^2_{1,l}},
\label{27}
\end{eqnarray}
where $\tilde{\omega}_{1,l}$ is given by Eq. (\ref{26}). The corresponding wave-function is given by
\begin{equation}
\psi_{1,l} (\rho)=\rho^{\sqrt{\frac{l^2}{\alpha^2}+m^2\,\omega^2\,b^2}}\,e^{-\frac{1}{2}\,(2\,c_1+\rho)\,\rho}\,(c_0+c_1\,\rho),
\label{28}
\end{equation}
where
\begin{equation}
c_1=\frac{1}{\sqrt{\frac{3}{2}+\sqrt{\frac{l^2}{\alpha^2}+m^2\,\omega^2\,b^2}}}.
\label{29}
\end{equation}

We discuss below few case of the above relativistic energy eigenvalue Eq. (\ref{24}) of the relativistic system.

\vspace{0.2cm}
{\bf Case 1} : If one choose $b \rightarrow 0$, $a \rightarrow 1$, that is, the Klein-Gordon oscillator subject to a linear scalar potential in the cosmic string space-time in the Kaluza-Klein theory.
\vspace{0.2cm}

The energy eigenvalue Eq. (\ref{24}) becomes
\begin{equation}
E^{2}_{n,l}=k^2+q^2+m^2+2\,m\,\omega+2\,\tilde{\omega}_{n,l}\,(n+1+\frac{|l|}{\alpha})-\frac{m^2\,k^{2}_{L}}{\tilde{\omega}^2_{n,l}}.
\label{30}
\end{equation}
Equation (\ref{30}) is the energy eigenvalue of the Klein-Gordon oscillator field in the cosmic string background within the Kaluza-Klein theory subject to a linear scalar potential. Note that for $k=0=q$, the energy eigenvalue Eq. (\ref{30}) reduces to the result obtained in \cite{aa49}. Thus the relativistic energy eigenvalue Eq. (\ref{30}) is the generalised result in five-dimensions in comparison to those obtained in (1+2)-dimensional case in  \cite{aa49}.

Therefore, the bound state energy eigenvalue for $n=1$ is given by
\begin{equation}
E^{2}_{1,l}=k^2+q^2+m^2+2\,m\,\omega+2\,\tilde{\omega}_{1,l}\,(n+1+\frac{|l|}{\alpha})-\frac{m^2\,k^{2}_{L 1}}{\tilde{\omega}^2_{1,l}}.
\label{31}
\end{equation}
And the corresponding wave function is
\begin{equation}
\psi_{1,l} (\rho)=\rho^{\frac{|l|}{\alpha}}\,e^{-\frac{1}{2}\,(2\,c_1+\rho)\,\rho}\,(c_0+c_1\,\rho),
\label{32}
\end{equation}
where
\begin{eqnarray}
&&c_1=\frac{1}{\sqrt{\frac{3}{2}+\frac{|l|}{\alpha}}},\nonumber\\
&&\tilde{\omega}_{1,l}=[m^2\,k^2_{L 1}\,(\frac{3}{2}+\frac{|l|}{\alpha})]^{\frac{1}{3}}.
\label{33}
\end{eqnarray}

\vspace{0.2cm}
{\bf Case 2} : If one choose $k_{L} \rightarrow 0$, that is, without any scalar potential in the relativistic system. 
\vspace{0.2cm}

From Eq. (\ref{11}), we obtain the following equation
\begin{equation}
\psi''(r)+\frac{1}{r}\,\psi'(r)+[\lambda-m^2\,\omega^2\,a^2\,r^2-\frac{\xi^2_{s}}{r^2}]\,\psi(r)=0,
\label{34}
\end{equation}
Introducing a new variable $s=m\,\omega\,a\,r^2$ into the above equation, we obtain \cite{bb49}
\begin{equation}
\psi''(s)+\frac{1}{s}\,\psi' (s)+\frac{1}{s^2}\,(-\xi_1\,s^2+\xi_2\,s-\xi_3)\,\psi (s)=0,
\label{35}
\end{equation}
where
\begin{equation}
\xi_1=\frac{1}{4}\quad,\quad \xi_2=\frac{\lambda}{4\,m\,\omega\,a}\quad,\quad \xi_3=\frac{\xi^2_{s}}{4}.
\label{36}
\end{equation}
Compairing the above Eq. (\ref{35}) with (\ref{A.1}) in appendix A, we get
\begin{eqnarray}
&&\alpha_1=1,\quad \alpha_2=0,\quad \alpha_3=0,\quad \alpha_4=0,\quad \alpha_5=0,\nonumber\\
&&\alpha_6=\xi_1,\quad \alpha_7=-\xi_2,\quad \alpha_8=\xi_3,\quad \alpha_9=\xi_1,\nonumber\\
&&\alpha_{10}=1+2\,\sqrt{\xi_3},\quad \alpha_{11}=2\,\sqrt{\xi_1},\quad \alpha_{12}=\sqrt{\xi_3},\quad \alpha_{13}=-\sqrt{\xi_1}.
\label{37}
\end{eqnarray}
The energy eigenvalue using (\ref{36})--(\ref{37}) into the Eq.(\ref{A.8}) in appendix A is given by
\begin{eqnarray}
&&(2\,n+1)\,\sqrt{\xi_1}-\xi_2+2\,\sqrt{\xi_1\,\xi_3}=0\nonumber\\\Rightarrow
E^{2}_{n,l}&=&k^2+q^2+m^2+2\,m^2\,\omega^2\,a\,b\nonumber\\
&+&2\,m\,\omega\,a\,(2\,n+2+\sqrt{\frac{l^2}{\alpha^2}+m^2\,\omega^2\,b^2}),
\label{38}
\end{eqnarray}
where $n=0,1,2,3,....$.

If one choose $b \rightarrow 0$, $a \rightarrow 1$, that is, the Klein-Gordon oscillator in the cosmic string without potential in the Kaluza-Klein theory. Then, the energy eigenvalue (\ref{38}) becomes
\begin{equation}
E^{2}_{n,l}=k^2+q^2+m^2+4\,m\,\omega\,(n+1+\frac{|l|}{2\,\alpha})
\label{39}
\end{equation}
which is similar to the result obtained in \cite{aa6} (see Eq. (12) in \cite{aa6}). We can see that the Eq. (\ref{38}) is the extended energy eigenvalue of the generalized Klein-Gordon oscillator in the cosmic string space-time without potential in the Kaluza-Klein theory.

\vspace{0.2cm}
{\bf Case 3} : If one choose $b \rightarrow 0$, $a \rightarrow 0$, the Klein-Gordon particle subject to linear scalar potential in a cosmic string background in a Kaluza-Klein theory.
\vspace{0.2cm}

In that case, the Eq. (\ref{11}) becomes
\begin{equation}
\psi''(r)+\frac{1}{r}\,\psi'(r)+[E^2-m^2-k^2-q^2-k^2_{L}\,r^2-\frac{l^2}{\alpha^2\,r^2}]\,\psi(r)=0.
\label{40}
\end{equation}
Transforming $s=K_{L}\,r^2$ into the above equation, we obtain the following differential equation \cite{bb49}
\begin{equation}
\psi''(s)+\frac{1}{s}\,\psi'(s)+\frac{1}{s^2}\,[\frac{E^2-m^2-k^2-q^2}{4\,k_{L}}\,s-\frac{1}{4}\,s^2-\frac{l^2}{4\,\alpha^2}]\,\psi(s)=0.
\label{41}
\end{equation}
As done earlier, we obtain the following energy eigenvalue of the system 
\begin{equation}
E_{n,l}=\pm\,\sqrt{k^2+q^2+m^2+2\,k_{L}\,(2\,n+1+\frac{|l|}{\alpha})}.
\label{42}
\end{equation}
Equation (\ref{42}) is the relativistic energy eigenvalue of a scalar particle in the cosmic string space-time in the Kaluza-Klein theory with a linear scalar potential.

\section{Generalized Klein-Gordon oscillator in the magnetic cosmic string space-time with a linear scalar potential in the Kaluza-Klein theory}

In this section, we investigate analogue effect to the Aharonov-Bohm effect for bound states \cite{aa6,bb28,bb48,bb39, bb40,bb50} solution of a scalar particle. By using the Kaluza-Klein theory, we introduce a magnetic flux through the line element of the cosmic string and thus write the generalized Klein-Gordon oscillator equation in five-dimensional space-time subject to a linear scalar potential. This kind of potential has great reserach interested and used for the confinement of quark models and in other branches of physics discussed earlier.

Let us consider the quantum dynamics of a scalar particle moving in the magnetic cosmic string background. In the Kaluza-Klein theory \cite{bb29}, the corresponding metric with the magnetic quantum flux $\Phi$ passing along the symmetry axis of the string assumes the following form
\begin{equation}
ds^2=-dt^2+dr^2+\alpha^2\,r^2\,d\phi^2+dz^2+(dx+\frac{\Phi}{2\,\pi}\,d\phi)^2
\label{43}
\end{equation}
with cylindrical coordinates are used. The quantum dynamics is described by the Eq. (\ref{2}) with the following change in the inverse matrix tensor $g^{\mu\nu}$,
\begin{equation}
g^{\mu\nu}=\left (\begin{array}{lllll}
-1 & 0 & \quad 0 & 0 & \quad 0 \\
\quad 0 & 1 & \quad 0 & 0 & \quad 0 \\
\quad 0 & 0 & \quad \frac{1}{\alpha^2\,r^2} & 0 & -\frac{\Phi}{2\,\pi\,\alpha^2\,r^2} \\
\quad 0 & 0 & \quad 0 & 1 & \quad 0 \\
\quad 0 & 0 & -\frac{\Phi}{2\,\pi\,\alpha^2\,r^2} & 0 & 1+\frac{\Phi^2}{4\,\pi^2\,\alpha^2\,r^2}
\end{array} \right).
\label{44}
\end{equation}
By considering the line element (\ref{43}) into the Eq. (\ref{5}), we obtain the following differential equation :
\begin{eqnarray}
&&[-\partial_{t}^2+\partial_{r}^2+\frac{1}{r}\,\partial_{r}+\frac{1}{\alpha^2\,r^2}\,(\partial_{\phi}-\frac{\Phi}{2\,\pi}\,\partial_{x})^2+\partial_{z}^2+\partial_{x}^2-m\,\omega\,(f' (r)+\frac{f(r)}{r})\nonumber\\
&&-m^2\,\omega^2\,f^{2}(r)-(m + S(r))^2]\,\Psi (r)=0.
\label{45}
\end{eqnarray}
Since the space-time is independent of $t, \phi, z, x$, substituting the ansatz Eq. (\ref{8}) into the Eq. (\ref{45}), we get the following equation :
\begin{eqnarray}
&&\psi ''(r)+\frac{1}{r}\,\psi'(r)+[E^2-k^2-q^2-\frac{l^2_{eff}}{r^2}-m\,\omega\,(f'(r)+\frac{f(r)}{r})\nonumber\\
&-&m^2\,\omega^2\,f^{2}(r)-(m+S (r))^2]\,\psi (r)=0,
\label{46}
\end{eqnarray}
where
\begin{equation}
l_{eff}=\frac{1}{\alpha}\,(l-\frac{q\,\Phi}{2\,\pi})
\label{47}
\end{equation}
is called the effective angular quantum number which depends on the cosmic string parameter as well as the magnetic quantum flux. 

Substituting the function Eq. (\ref{function}) into the Eq. (\ref{46}) and using the scalar potential Eq. (\ref{10}), we obtain the following equation:
\begin{equation}
\psi''(r)+\frac{1}{r}\,\psi'(r)+[\lambda-\tilde{\omega}^2\,r^2-\frac{\chi^2_{eff}}{r^2}-2\,m\,k_{L}\,r]\,\psi(r)=0,
\label{48}
\end{equation}
where
\begin{eqnarray}
&&\lambda=E^2-k^2-q^2-m^2-2\,m\,\omega\,a-2\,m^2\,\omega^2\,a\,b,\nonumber\\
&& \tilde{\omega}=\sqrt{m^2\,\omega^2\,a^2+k^{2}_{L}},\nonumber\\
&&\chi_{eff}=\sqrt{l^2_{eff}+m^2\,\omega^2\,b^2}.
\label{49}
\end{eqnarray}
Transforming $\rho=\sqrt{\tilde{\omega}}\,r$ into the Eq. (\ref{48}), we get
\begin{equation}
\psi''(\rho)+\frac{1}{\rho}\,\psi'(\rho)+[\lambda_0-\rho^2-\frac{\chi^2_{eff}}{\rho^2}-\theta\,\rho]\,\psi (\rho)=0,
\label{50}
\end{equation}
where $\lambda_0, \theta$ are in Eq. (\ref{aa}).

As similar to the technique done earlier, suppose the solution to the Eq. (\ref{50}) is
\begin{equation}
\psi (\rho)=\rho^{\chi_{eff}}\,e^{-\frac{1}{2}\,(\theta+\rho)\,\rho}\,H (\rho).
\label{51}
\end{equation}
Substituting Eq. (\ref{51}) into the Eq. (\ref{50}), we get the biconfluent Heun's differential equation \cite{aa30,aa49,aa50,bb41,bb42,ff2,ff3} form:
\begin{equation}
H''(\rho)+[\frac{\bar \gamma}{\rho}-\theta-2\,\rho]\,H'(\rho)+[-\frac{\bar Q}{r}+{\bar \Theta}]\,H (\rho)=0,
\label{52}
\end{equation}
where
\begin{eqnarray}
&&{\bar \gamma}=(1+2\,\chi_{eff}),\nonumber\\
&&{\bar \Theta}=\lambda_0+\frac{\theta^2}{4}-2\,(1+\chi_{eff}),\nonumber\\
&&{\bar Q}=\frac{\theta}{2}\,(1+2\,\chi_{eff}).
\label{53}
\end{eqnarray}

Substituting Eq. (\ref{19}) into the Eq. (\ref{52}), we get the following recurrernce relation for the coefficients:
\begin{equation}
c_{n+2}=\frac{1}{(n+2)(n+1+\gamma)}\,[\{{\bar Q}+\theta\,(n+1) \}\,c_{n+1}-({\bar \Theta}-2\,n)\,c_{n}].
\label{54}
\end{equation}
And the various coefficients are
\begin{equation}
c_1=\frac{\bar Q}{\bar \gamma}\,c_0,\quad c_2=\frac{1}{2\,(1+\gamma)}\,[({\bar Q}+\theta)\,c_{1}-{\bar \Theta}\,c_{0}].
\label{55}
\end{equation}
A polynomial form of degree $n$ for the function $H (r)$ is achieved when we impose requirement that the series solution terminates. For this, we must have \cite{aa30,aa50,ff2,ff3,bb41,bb44,bb45,bb46,bb47}
\begin{equation}
{\bar \Theta}=2\,n.
\label{56}
\end{equation}
And
\begin{equation}
c_{n+1}=0,\quad (n=1,2,3,4,....).
\label{57}
\end{equation}
Using the condition ${\bar \Theta}=2\,n$, we obtain the following energy eigenvalue $E_{n,l}$:
\begin{eqnarray}
E^{2}_{n,l}&=&k^2+q^2+m^2+2\,m\,\omega\,a+2\,m^2\,\omega^2\,a\,b-\frac{m^2\,k^{2}_{L}}{\tilde{\omega}^2}\nonumber\\
&+&2\,\tilde{\omega}\,(n+1+\sqrt{\frac{1}{\alpha^2}\,(l-\frac{q\,\Phi}{2\,\pi})^2+m^2\,\omega^2\,b^2}).
\label{58}
\end{eqnarray}
Equation (\ref{58}) is the energy eigenvalue associated with $n^{th}$ radial modes for the generalized Klein-Gordon oscillator in the magnetic cosmic string with a linear scalar potential in the Kaluza-Klein theory. We can see in comparision to the case without magnetic flux as obtained earlier by Eq. (\ref{24}), the angular quantum number $l$ is shifted, $l \rightarrow l_{eff}=\frac{1}{\alpha}\,(l-\frac{q\,\Phi}{2\,\pi})$, an effective angular quantum number. We see that the relativistic energy eigenvalue Eq. (\ref{58}) depends on the Aharonov-Bohm geometric quantum phase \cite{bb50}. Thus, we have that $E_{n,{\bar l}} (\Phi+\Phi_0)=E_{n,{\bar l}\pm \tau} (\Phi)$ where, $\Phi_0=\mp\,\frac{2\,\pi\,\alpha}{q}\,\tau$ with $\tau=1,2,3,...$ and ${\bar l}=\frac{l}{\alpha}$. This dependence of the relativistic energy eigenvalue on the geometric quantum phase gives rise to the analogoue effects to the Aharonov-Bohm effect for bound states \cite{bb28,bb48,bb39,bb40,bb50}.

The corresponding wave-function is given by
\begin{equation}
\psi_{n,l} (\rho)=\rho^{\sqrt{\frac{(l-\frac{q\,\Phi}{2\,\pi})^2}{\alpha^2}+m^2\,\omega^2\,b^2}}\,e^{-\frac{1}{2}\,(\theta+\rho)\,\rho}\,H (\rho).
\label{59}
\end{equation}

Now, we impose the additional condition $c_{n+1}=0$ in the above eigenvalue problem to obtain the individual energy levels and corresponding wave-function one by one as done in \cite{aa30,aa50,bb44,bb45}. As example, for $n=1$, we have $c_2=0$ which implies
\begin{eqnarray}
&&c_1=\frac{2}{{\bar Q}+{\bar \beta}}\,c_0 \Rightarrow \frac{{\bar Q}}{{\bar \gamma}}=\frac{2}{{\bar Q}+{\bar \beta}}\nonumber\\\Rightarrow
&&B^2_{2}=\frac{1}{1+A_2} \Rightarrow \tilde{\omega}_{1,l}=[m^2\,k^2_{L 1}\,(\frac{3}{2}+\sqrt{\frac{(l-\frac{q\,\Phi}{2\,\pi})^2}{\alpha^2}+m^2\,\omega^2\,b^2})]^{\frac{1}{3}}.
\label{60}
\end{eqnarray}
Therefore, the ground state energy level for $n=1$ is given by
\begin{eqnarray}
&&E^{2}_{1,l}=k^2+q^2+m^2+2\,m\,\omega\,a+2\,m^2\,\omega^2\,a\,b\nonumber\\
&&+2\,\tilde{\omega}_{1,l}\,(n+1+\sqrt{\frac{(l-\frac{q\,\Phi}{2\,\pi})^2}{\alpha^2}+m^2\,\omega^2\,b^2})-\frac{m^2\,k^{2}_{L 1}}{\tilde{\omega}^2_{1,l}},
\label{61}
\end{eqnarray}
where $\tilde{\omega}_{1,l}$ is given by Eq. (\ref{60}). The corresponding wave-function is
\begin{equation}
\psi_{1,l} (\rho)=\rho^{\sqrt{\frac{(l-\frac{q\,\Phi}{2\,\pi})^2}{\alpha^2}+m^2\,\omega^2\,b^2}}\,e^{-\frac{1}{2}\,(2\,c_1+\rho)\,\rho}\,(c_0+c_1\,\rho),
\label{62}
\end{equation}
where
\begin{equation}
c_1=\frac{1}{\sqrt{\frac{3}{2}+\sqrt{\frac{(l-\frac{q\,\Phi}{2\,\pi})^2}{\alpha^2}+m^2\,\omega^2\,b^2}}}.
\label{63}
\end{equation}

Below we discuss few cases of the above energy eigenvalue of the relativistic system considered in this section.

\vspace{0.2cm}
{\bf Case 1} : If one choose $b \rightarrow 0$, $a \rightarrow 1$.
\vspace{0.2cm}

In that case, the energy eigenvalue Eq. (\ref{58}) becomes
\begin{equation}
E^{2}_{n,l}=k^2+q^2+m^2+2\,m\,\omega+2\,\tilde{\omega}_{n,l}\,(n+1+\frac{|l-\frac{q\,\Phi}{2\,\pi}|}{\alpha})-\frac{m^2\,k^{2}_{L}}{\tilde{\omega}^2_{n,l}}.
\label{64}
\end{equation}
Equation (\ref{64}) is the energy eigenvalue associated with $n^{th}$ radial modes for the Klein-Gordon oscillator in the magnetic cosmic string background with a linear scalar potential in the Kaluza-Klein theory.

The bound state energy level for $n=1$ is given by
\begin{eqnarray}
&&E^{2}_{1,l}=k^2+q^2+m^2+2\,m\,\omega\,a\nonumber\\
&&+2\,\tilde{\omega}_{1,l}\,(n+1+\frac{|l-\frac{q\,\Phi}{2\,\pi}|}{\alpha})-\frac{m^2\,k^{2}_{L 1}}{\tilde{\omega}^2_{1,l}}.
\label{65}
\end{eqnarray}
And the corresponding wave-function is 
\begin{equation}
\psi_{1,l} (\rho)=\rho^{\frac{|l-\frac{q\,\Phi}{2\,\pi}|}{\alpha}}\,e^{-\frac{1}{2}\,(2\,c_1+\rho)\,\rho}\,(c_0+c_1\,\rho),
\label{66}
\end{equation}
where
\begin{eqnarray}
&&c_1=\frac{1}{\sqrt{\frac{3}{2}+\frac{|l-\frac{q\,\Phi}{2\,\pi}|}{\alpha}}},\nonumber\\
&&\tilde{\omega}_{1,l}=[m^2\,k^2_{L 1}\,(\frac{3}{2}+\frac{|l-\frac{q\,\Phi}{2\,\pi}|}{\alpha})]^{\frac{1}{3}}.
\label{67}
\end{eqnarray}

\vspace{0.1cm}
{\bf Case 2} : If one choose $k_{L} \rightarrow 0$, that is, without any linear scalar potential into the considered system. 
\vspace{0.1cm}

In that case, from Eq. (\ref{48}) we have the following equation
\begin{equation}
\psi''(r)+\frac{1}{r}\,\psi'(r)+[\lambda-m^2\,\omega^2\,a^2\,r^2-\frac{\chi^2_{eff}}{r^2}]\,\psi(r)=0
\label{68}
\end{equation}
which can be transformed to the following equation \cite{bb49}
\begin{equation}
\psi''(s)+\frac{1}{s}\,\psi' (s)+\frac{1}{s^2}\,(-\xi_1\,s^2+\xi_2\,s-\xi_3)\,\psi (s)=0,
\label{69}
\end{equation}
where
\begin{equation}
\xi_1=\frac{1}{4}\quad,\quad \xi_2=\frac{\lambda}{4\,m\,\omega\,a}\quad,\quad \xi_3=\frac{\chi^2_{eff}}{4}.
\label{70}
\end{equation}

The energy eigenvalue is given by
\begin{eqnarray}
E^{2}_{n,l}&=&k^2+2\,m^2\,\omega^2\,a\,b+q^2+m^2\nonumber\\
&+&2\,m\,\omega\,a\,(2\,n+2+\sqrt{\frac{(l-\frac{q\,\Phi}{2\,\pi})^2}{\alpha^2}+m^2\,\omega^2\,b^2}),
\label{71}
\end{eqnarray}
where $n=0,1,2,3,...$. Equation (\ref{71}) is the energy eigenvalue associated with $n^{th}$ radial modes for the generalized Klein-Gordon oscillator in the magnetic cosmic string without potential in the Kaluza-Klein theory.

If one choose $b \rightarrow 0$, $a \rightarrow 1$, that is, the Klein-Gordon oscillator in the magnetic cosmic string background in the Kaluza-Klein theory, the energy eigenvalue Eq. (\ref{71}) becomes
\begin{equation}
E^{2}_{n,l}=k^2+q^2+m^2+4\,m\,\omega\,(n+1+\frac{|l-\frac{q\,\Phi}{2\,\pi}|}{2\,\alpha})
\label{72}
\end{equation}
which is similar to the result obtained in \cite{aa6} (see Eq. (25) in \cite{aa6}). Thus the Eq. (\ref{71}) is the extended energy eigenvalue of the generalized Klein-Gordon oscillator in the magnetic cosmic string background without potential in the Kaluza-Klein theory.

\vspace{0.2cm}
{\bf Case 3} : If one choose $b \rightarrow 0$, $a \rightarrow 0$, the Klein-Gordon particle subject to a linear scalar potential into the relativistic system.
\vspace{0.2cm}

In that case, the Eq. (\ref{61}) becomes
\begin{equation}
\psi''(r)+\frac{1}{r}\,\psi'(r)+[E^2-m^2-k^2-q^2-k^2_{L}\,r^2-\frac{l^2_{eff}}{r^2}]\,\psi(r)=0.
\label{73}
\end{equation}
Transforming $s=k_{L}\,r^2$ into the above equation, we obtain \cite{bb49}
\begin{equation}
\psi''(s)+\frac{1}{s}\,\psi'(s)+\frac{1}{s^2}\,[\frac{E^2-m^2-k^2-q^2}{4\,k_{L}}\,s-\frac{1}{4}\,s^2-\frac{l^2_{eff}}{4}]\,\psi(s)=0.
\label{74}
\end{equation}
The energy eigenvalue of the system is
\begin{equation}
E^{2}_{n,l}=k^2+q^2+m^2+2\,k_{L}\,(2\,n+1+\frac{|l-\frac{q\,\Phi}{2\,\pi}|}{\alpha}),
\label{75}
\end{equation}
where $n=0,1,2,3,....$. Equation (\ref{75}) is the energy eigenvalue of a scalar field in the magnetic cosmic string background in the Kaluza-Klein theory with a linear scalar potential.

In all the above cases, we can see that there is an effective angular quantum number, $l \rightarrow l_{eff}=\frac{1}{\alpha}\,(l-\frac{q\,\Phi}{2\,\pi})$. Therefore, the above relativistic energy eigenvalue depends on the geometric quantum phase \cite{bb50}. Thus, we have that $E_{n,{\bar l}} (\Phi+\Phi_0)=E_{n,{\bar l} \pm \tau} (\Phi)$ where, $\Phi_0=\mp\,\frac{2\,\pi\,\alpha}{q}\,\tau$ with $\tau=1,2,3,....$ and ${\bar l}=\frac{l}{\alpha}$. This dependence of the relativistic energy eigenvalue on the geometric quantum phase gives rise to the analogoue effect to the Aharonov-Bohm effect for bound states \cite{aa6,bb28,bb48,bb39,bb40,bb50}.

Formula (\ref{47}) suggests that, when the particle circles the string, the wave-function changes according to
\begin{equation}
\Psi\rightarrow \Psi'=e^{2\,i\,\pi\,l_{eff}}\,\Psi=e^{i\,\frac{2\,\pi}{\alpha}\,(l-\frac{q\,\Phi}{2\,\pi})}\,\Psi.
\label{76}
\end{equation}
An immediate consequence of Eq. (\ref{47}) is that the angular momentum operator may be redefined as
\begin{equation}
\hat{l}_{eff}=-\frac{i}{\alpha}\,(\partial_{\phi}-i\,\frac{q\,\Phi}{2\,\pi}),
\label{77}
\end{equation}
where the additional term, $-\frac{q\,\Phi}{2\,\pi\,\alpha}$, takes into account the Aharonov-Bohm magnetic quantum flux $\Phi$.

\vspace{0.5cm}
\section{Conclusions}

In Ref. \cite{aa6}, authors studied the relativistic scalar particle in the cosmic string, magnetic cosmic string and cosmic dispiration background in the Kaluza-Klein theory. They solved the Klein-Gordon oscillator without any potential and obtained the relativistic energy eigenvalue and eigenfunction. In Ref. \cite{aa49}, the relativistic quantum dynamics of a scalar particle in (1+2)-dimensional space-time with topological defects subject to a linear scalar potential were studied. They solved the Klein-Gordon oscillator in the considered framework and obtained the relativistic energy eigenvalue and wave-function.

In this work, we have investigated the relativistic quantum dynamics of a scalar particle interacting with gravitational field produced by topological defects via the generalized Klein-Gordon oscillator in the cosmic string and magnetic cosmic string space-time within the Kaluza-Klein theory subject to a linear scalar potential. We have determined the manner in which the non-trivial topology due to topological defects and the magnetic quantum flux modifies the energy spectrum and wave-function of the relativistic system. We have studied the quantum dynamics of a scalar particle interacting with an external field sources, by using the five-dimensional version of the General Relativity. The quantum dynamics in the usual as well as magnetic cosmic string cases allow us to obtain the energy eigenvalue and wave-functions depending on the external parameters characterizing the background space-time, a result known by gravitational analogue of the well studied Aharonov-Bohm effect.

In {\it section 2}, we have studied the relativistic quantum dynamics of a scalar particle in the background of cosmic string space-time in the Kaluza-Klein theory with a linear scalar potential. In this study, we have considered the Cornell-type potential form function $f(r)=a\,r+\frac{b}{r}$ \cite{cc9,bb41} and a linear scalar potential $S(r)=k_{L}\,r$ where, $k_{L}$ is the linear confining parameter. Then, we have solved the generalized Klein-Gordon oscillator subject to this potential in the cosmic string background space-time in the Kaluza-Klein theory. We have obtained the bound states energy eigenvalue Eq. (\ref{24}) and corresponding wave-function Eq. (\ref{25}). By imposing the additional recurrence condition $c_{n+1}=0$ in the eiegnvalue problem, for example, $n=1$, we have evaluated the ground state energy level  Eq. (\ref{27}) and corresponding wave-function Eqs. (\ref{28})-(\ref{29}), respectively. We have seen that gravitational field produced by topological defects and the scalar potential modifies the energy spectrum. Furthermore, we have discussed three cases {\it 1-3} and seen that the energy eigenvalue in very special case reduces to the result obtained in \cite{aa49} (for case {\bf 1}) and in \cite{aa6} (for case {\bf 2}).

In {\it section 3}, we have studied the relativistic quantum dynamics of a scalar particle in the magnetic cosmic string within the Kaluza-Klein theory with a linear scalar potential. We have chosen the same function $f(r)=a\,r+\frac{b}{r}$ considered earlier and the linear scalar potential $S(r)=k_{L}\,r$ into the relativistic system. We have derived the radial wave-equation of the generalized Klein-Gordon oscillator by choosing this function $f(r)$ subject to this potential in the magnetic cosmic string in the Kaluza-Klein theory. We have obtained the bound states energy eigenvalue Eq. (\ref{58}) and corresponding wave-function Eq. (\ref{59}). By imposing the additional condition $c_{n+1}=0$, for example $n=1$, we have obtained the ground state energy level and eigenfunction. In addition, we have discussed few cases {\it 1-3} and obtained the relativistic energy eigenvalue. We have seen in case {\bf 2} that the energy eigenvalue reduces to the result obtained in \cite{aa6} in very special case. In this section, we have seen that the relativistic energy eigenvalue obtained here depends on the geometric quantum phase \cite{bb50}. Thus, we have that $E_{n,{\bar l}} (\Phi+\Phi_0)=E_{n,{\bar l} \pm \tau} (\Phi)$ where, $\Phi_0=\mp\,\frac{2\,\pi\,\alpha}{q}\,\tau$ with $\tau=1,2,3,...$ and ${\bar l}=\frac{l}{\alpha}$. This dependence of the relativistic energy eigenvalue on the geometric quantum phase $\Phi$ gives rise to the analogoue effect to the Aharonov-Bohm effect for bound states \cite{bb28,bb48,bb39,bb40,bb50}. Besides, even though there is no direct intractions between the particle and external fields, influence on the energy eigenvalue is due to the internal magnetic quantum flux $\Phi$ yielding an effective angular quantum number $l \rightarrow l_{eff}=\frac{1}{\alpha}\,(l-\frac{q\,\Phi}{2\,\pi})$. For $\alpha \rightarrow 1$, the change in angular quantum number $\bigtriangleup\,l=l-l_{eff}$ is directly proportional to the magnetic quantum flux $\Phi$. So we have shown some results which are in addition to the previous results obtained in \cite{aa6,aa49} present many interesting effects. This is the fundamental subject in physics and the connection between these theories (gravitation and quantum mechanics) are not well understood.

\section*{Appendix A : Brief review of the Nikiforov-Uvarov (NU) method}

\setcounter{equation}{0}
\renewcommand{\theequation}{A.\arabic{equation}}

The Nikiforov-Uvarov method is helpful in order to find eigenvalues and eigenfunctions of the Schr\"{o}dinger like equation, as well as other second-order differential equations of physical interest. According to this method, the eigenfunctions of a second-order differential equation \cite{bb49}
\begin{equation}
\frac{d^2 \psi (s)}{ds^2}+\frac{(\alpha_1-\alpha_2\,s)}{s\,(1-\alpha_3\,s)}\,\frac{d \psi (s)}{ds}+\frac{(-\xi_1\,s^2+\xi_2\,s-\xi_3)}{s^2\,(1-\alpha_3\,s)^2}\,\psi (s)=0.
\label{A.1}
\end{equation}
are given by 
\begin{equation}
\psi (s)=s^{\alpha_{12}}\,(1-\alpha_3\,s)^{-\alpha_{12}-\frac{\alpha_{13}}{\alpha_3}}\,P^{(\alpha_{10}-1,\frac{\alpha_{11}}{\alpha_3}-\alpha_{10}-1)}_{n}\,(1-2\,\alpha_3\,s).
\label{A.2}
\end{equation}
And that the energy eigenvalues equation
\begin{eqnarray}
&&\alpha_2\,n-(2\,n+1)\,\alpha_5+(2\,n+1)\,(\sqrt{\alpha_9}+\alpha_3\,\sqrt{\alpha_8})+n\,(n-1)\,\alpha_3+\alpha_7\nonumber\\
&&+2\,\alpha_3\,\alpha_8+2\,\sqrt{\alpha_8\,\alpha_9}=0.
\label{A.3}
\end{eqnarray}
The parameters $\alpha_4,\ldots,\alpha_{13}$ are obatined from the six parameters $\alpha_1,\ldots,\alpha_3$ and $\xi_1,\ldots,\xi_3$ as follows:
\begin{eqnarray}
&&\alpha_4=\frac{1}{2}\,(1-\alpha_1)\quad,\quad \alpha_5=\frac{1}{2}\,(\alpha_2-2\,\alpha_3),\nonumber\\
&&\alpha_6=\alpha^2_{5}+\xi_1\quad,\quad \alpha_7=2\,\alpha_4\,\alpha_{5}-\xi_2,\nonumber\\
&&\alpha_8=\alpha^2_{4}+\xi_3\quad,\quad \alpha_9=\alpha_6+\alpha_3\,\alpha_7+\alpha^{2}_3\,\alpha_8,\nonumber\\
&&\alpha_{10}=\alpha_1+2\,\alpha_4+2\,\sqrt{\alpha_8}\quad,\quad \alpha_{11}=\alpha_2-2\,\alpha_5+2\,(\sqrt{\alpha_9}+\alpha_3\,\sqrt{\alpha_8}),\nonumber\\
&&\alpha_{12}=\alpha_4+\sqrt{\alpha_8}\quad,\quad \alpha_{13}=\alpha_5-(\sqrt{\alpha_9}+\alpha_3\,\sqrt{\alpha_8}).
\label{A.4}
\end{eqnarray}

A special case where $\alpha_3=0$, as in our case, we find
\begin{equation}
\lim_{\alpha_3\rightarrow 0} P^{(\alpha_{10}-1,\frac{\alpha_{11}}{\alpha_3}-\alpha_{10}-1)}_{n}\,(1-2\,\alpha_3\,s)=L^{\alpha_{10}-1}_{n} (\alpha_{11}\,s),
\label{A.5}
\end{equation}
and 
\begin{equation}
\lim_{\alpha_3\rightarrow 0} (1-\alpha_3\,s)^{-\alpha_{12}-\frac{\alpha_{13}}{\alpha_3}}=e^{\alpha_{13}\,s}.
\label{A.6}
\end{equation}
Therefore the wave-function from (\ref{A.2}) becomes
\begin{equation}
\psi (s)=s^{\alpha_{12}}\,e^{\alpha_{13}\,s}\,L^{\alpha_{10}-1}_{n} (\alpha_{11}\,s),
\label{A.7}
\end{equation}
where $L^{(\alpha)}_{n} (x)$ denotes the generalized Laguerre polynomial. 

The energy eigenvalues equation reduces to 
\begin{equation}
n\,\alpha_2-(2\,n+1)\,\alpha_5+(2\,n+1)\,\sqrt{\alpha_9}+\alpha_7+2\,\sqrt{\alpha_8\,\alpha_9}=0.
\label{A.8}
\end{equation}
Noted that the simple Laguerre polynomial is the special case $\alpha=0$ of the generalized Laguerre polynomila:
\begin{equation}
L^{(0)}_{n} (x)=L_{n} (x).
\label{A.9}
\end{equation}

\section*{Acknowledgement}

Author acknowledged the anonymous kind referee(s) for their valuable comments and suggestions which have improved the present paper.

\section*{Data Availability}

There is no data associated with this paper or no data have been used to prepare this.

\end{document}